\documentclass[journal,comsoc]{IEEEtran}
\usepackage[T1]{fontenc}
\usepackage{amsmath,amsthm,dsfont,tabularx}
\usepackage[cmintegrals]{newtxmath}
\usepackage[nocompress]{cite}
\usepackage[utf8]{inputenc}
\usepackage[english]{babel}
\usepackage[dvips]{graphicx}
\usepackage{subcaption}
\usepackage{algorithm,algorithmic}
\usepackage{caption}
\usepackage{bm}
\usepackage[noblocks]{authblk}
\usepackage{color}
\usepackage{multirow}
\usepackage{siunitx}
\usepackage{setspace}
\usepackage[dvipsnames]{xcolor}

\newtheorem{theorem}{Theorem}[section]

\floatname{Procedure}{algorithm}

\begin{document}
\title{Empirical Optimization on Post-Disaster Communication Restoration for Social Equality}

\author{Jianqing Liu*, Shangjia Dong, Thomas Morris
\thanks{*Corresponding Author}
\thanks{J. Liu and T. Morris are with the Department
of Electrical and Computer Engineering, University of Alabama in Huntsville, Huntsville,
AL 35899 USA e-mail: \{jianqing.liu, tommy.morris\}@uah.edu}
\thanks{S. Dong is with the Department
of Civil and Environmental Engineering, University of Delaware, Newark, 
DE 19716 USA e-mail: sjdong@udel.edu}
}
\maketitle

\begin{abstract}
Disasters are constant threats to humankind, and beyond losses in lives, they cause many implicit yet profound societal issues such as wealth disparity and digital divide. Among those recovery measures in the aftermath of disasters, restoring and improving communication services is of vital importance. Although existing works have proposed many architectural and protocol designs, none of them have taken human factors and social equality into consideration. Recent sociological studies have shown that people from marginalized groups (e.g., minority, low income, and poor education) are more vulnerable to communication outages. In this work, we take pioneering efforts in integrating human factors into an empirical optimization model to determine strategies for post-disaster communication restoration. We cast the design into a mix-integer non-linear programming problem, which is proven too complex to be solved. Through a suite of convex relaxations, we then develop heuristic algorithms to efficiently solve the transformed optimization problem. Based on a collected dataset, we further evaluate and demonstrate how our design will prioritize communication services for vulnerable people and promote social equality compared with an existing modeling benchmark. 
\end{abstract}


\IEEEpeerreviewmaketitle

\section{Introduction}
Communication service is critical in connecting people, and even more so in facing a disaster. From the deadly COVID-19 to forest fires and hurricanes, disasters have inflicted catastrophic impact on millions across nations. Disaster-disrupted services not only directly impact the health and economic well-being of our society but also exacerbate the societal disparities \cite{patel2020poverty}, particularly among socially vulnerable communities. In the immediate aftermath of disasters, people at the heart of disaster sites are in critical need of lifeline services, including transportation, water, electricity, and communication services. Failure to timely address these needs could cause further damages to communities and residents therein. 

In post disaster recovery, restoring or improving communication service is equally important as restoring clean water and electricity. This year's pandemic taught us this lesson in a vivid yet hard way: quarantined households rely on the Internet for work, school, and healthcare while those who cannot access reliable Internet are at severe disadvantages. An early news went viral that reported two elementary students use Taco-Bell's parking lot for free WiFi because they do not have Internet at home \cite{taco20}. Disasters like COVID-19 may not directly damage communication infrastructures, but they could lead to the surging demands on communication resources that overburden existing infrastructures. If no measures are taken, certain groups of people may receive inadequate communication services which would worsen the so-called ``digital divide'' and breed social inequality \cite{warschauer2004technology}. For example, during COVID-19 lock-downs, rural and low-income people are found to experience more hardship to Internet access than their urban and high-income peers \cite{peters2020community}. In fact, according to recent studies \cite{esmalian2021susceptibility,esmalian2020empirical,dong2020integrated}, people of a different race, ethnicity, and age, in different geographic regions, with different education levels present significantly distinct hardship or vulnerability to disasters.

In light of it, to precisely address personalized communication needs, the inclusion of human factors is of vital importance. Existing works \cite{deepak2019overview,manoj2007communication,ran2011considerations,lieser2019understanding,liu2018resource,ochoa2015human} on post-disaster communication restoration or service provisioning mainly focus on the development of ad hoc or long-range communication architectures and protocols for delivering information, especially urgent ones, to/from the disaster sites. Unfortunately, none of them have taken human factors into consideration. This deficit in the state-of-the-art will not only result in impracticality of their solutions but also exacerbate the risks in social equality.

In this work, we aim to bridge this gap by developing empirical analysis and optimization models for post-disaster communication service provisioning based on the critical needs of under-represented users. This work will shed light on how different groups of users (categorized by their sociodemographic characteristics) have different levels of needs for communication services, and how stakeholders (e.g., government, Internet service providers) could provide a more socially equitable communication services. To this end, there are three challenges to be addressed. (1) There lacks empirical understanding on users' vulnerabilities to communication disruptions and their service needs after real disaster events. (2) It is not well understood what a practical and usable modeling should be so as to deliver prioritized services to targeted groups. Unfortunately, conventional unitary modeling (e.g., coverage- or throughput-maximization) for post-disaster communication service provisioning does not fully reflect human needs. (3) It is computationally difficult to solve and analyze empirical models on large datasets, which is mostly attributed to the binary decision variables. While the solution to mix-integer optimization problems is widely seen, empirical optimizations pose a greater challenge due to the prohibitively large scale of the problem. The efforts in tackling the above challenges mark our contribution in this work, which are as follows.
\vspace{-.01in}
\begin{itemize}
  \item Based on an empirical household survey data collected in the aftermath of Hurricane Harvey in Harris County, we examine various subjective perceptions on communication services from different population groups (categorized based on their sociodemographic attributes).
  \item We propose an inclusive model for post-disaster communication restoration and improvement, which reflects users' specific communication needs with respect to (w.r.t.) latency and bandwidth, thus prioritizing services for socially under-represented users.
  \item We develop a heuristic algorithm based on spatial-temporal characteristics of the dataset to efficiently calculate near-optimal solutions from the large-scale optimization problem.
\end{itemize}
\vspace{-.01in}

The rest of the paper is organized as follows. Section II surveys the most recent literature on this topic. Section III describes the system model. The problem formulation is outlined in Section IV. We propose solution algorithms in Section V and present a benchmark scheme from recent literature. In Section VI, we conduct the performance evaluation and demonstrate simulation results. Section VII discusses the major takeaways and outlines future directions. Finally, the paper is concluded in Section VIII.

\section{Related Works}
The focus of this work is reminiscent of two similar studies in recent literature: (1) post-disaster communication restoration, and (2) user-centric (or social-aware) designs in communications and networking. In this section, we discuss related works following these theme lines and present how this work contributes to the state-of-the-art.

\subsubsection{Post-disaster communication restoration} Amid highly destructive disasters such as earthquakes and tsunami, local communication infrastructures could become absolutely non-functional \cite{tapolcai2020fast}. In light of it, several proposals have presented designs to temporarily establish communication capabilities to deliver emergency information. For instance, the infrastructure-less ad hoc network based on (unmanned) aerial vehicles \cite{liu2018resource,erdelj2017help}, ground vehicles \cite{li2017vehicle} and marine vehicles \cite{murphy2011use} is a popular approach to serve this purpose. Other solutions include, but are not limited to, leveraging satellite networks \cite{tani2017flexibility}, LTE/LTE-A networks \cite{deepak2019overview,casoni2015integration}, device-to-device (D2D) networks \cite{ali2016architecture}, and Internet-of-Things (IoT) sensory networks \cite{suri2018exploiting}. Most of these works only carry high-level studies (e.g., architectural design, feasibility analysis, and system implementation) and communication-specs oriented designs (e.g., to maximize network throughput, extend coverage/connectivity, or reduce latency). In recent studies, some works started to look into message differentiation and prioritization in post-disaster communications. Lieser et al. \cite{lieser2019understanding} investigated the timing patterns of different messages (e.g., SOS, ``I'm alive'') after disasters and provided insights on which messages should be prioritized at what time. Bhattacharjee et al. \cite{bhattacharjee2016best} designed a natural language processing (NLP) algorithm to filter emergency messages which are delivered using a developed priority-enhanced routing protocol. 

Nevertheless, none of the existing works have ever taken human factors into consideration. Sociological studies have shown that people of a different race, gender, age, income level and etc., suffer distinctly in the aftermath of disasters. For instance, Esmalian et al. \cite{esmalian2021susceptibility} showed that people with personal vehicles are less susceptible to the loss of communication services compared with those low-income ones who have no vehicles. Coleman et al. \cite{coleman2020equitable} unveiled that renters experience higher hardship and have lower tolerance towards communication disruptions. Failure to address the critical needs of different subpopulations will not only reduce the efficacy in post-disaster communication provisioning but most importantly, increase risks of social disparities.

\subsubsection{User-centric communication and networking designs} By far, there are numerous user-centric \cite{liu2016energy,li2020user,liu2020optimizing,yang2015incentive} or social-aware \cite{ahmed2017social,gao2012social} research works that involve human subjects in the communications, computing and networking designs. Since the number of the related works is too significant to exhaust, we only discuss the common characteristics in their modeling and present how this work differs from them. Basically, related studies develop models by assumably or artificially translating user demands, constraints, or social interactions into some quantitative values. For example, Liu et al. \cite{liu2020optimizing} transformed users' difficulty in accessing external battery power into weighting factors when they attempted to maximize users' energy efficiency in an LTE network. Yang et al. \cite{yang2015incentive} developed an incentive model for crowdsensing by subjectively mapping users' satisfaction/utility to a quantitative value. Gao et al. \cite{gao2012social} designed a relay selection mechanism in delay tolerant networks (DTN) by modeling users' physical contact traces in a bipartite graph. 

However, there are two major limitations in these works. (1) Simply mapping human-related factors into assumable values raises concerns on how honest and reflective the developed models are to practical scenarios. (2) There are little to none works that have directly integrated people's sociodemographic factors and needs into their design. For example, people of old age concerning their privacy are reluctant to share their devices as a relay; people of high income are not easily biased by payment-based incentives.

\subsubsection{How our work differs with and contributes to the state-of-the-art} To address the inefficacy of existing works along these two theme lines, we develop an empirical model based on household survey data in the aftermath of Hurricane Harvey. By empirically analyzing human factors and optimizing communication resources, this work will (1) shed light on high-fidelity modeling of user-centric designs (coined as technical contribution), and (2) offer targeted and socially equitable communication service provisioning in post disasters (coined as practical contribution).

\section{System Model}
We consider a geographic region ${\cal N}$ in which users experience degraded or loss of communication services after a disaster. Suppose the whole region can be divided into discrete areas of interest ${\cal N} = \{1,...,N\}$ according to communities, zip codes, etc. Suppose there are $K_{n}$ users in a specific area $n \in {\cal N}$ and they are denoted as ${\cal U}_{n} = \{u_{1, n},...,u_{K_{n}, n}\}$. Collectively, users in all discrete areas constitute the whole user set $\overrightarrow{\cal U} = \{{\cal U}_{1},...,{\cal U}_{N}\}$ in the region. We assume that every user $u_{k, n}$ can be characterized by a three-tuple attribute specified as $\pi = \langle\tau, \hat{\mathrm{r}}, \mathrm{d}\rangle$ (we drop the subscript for brevity), which respectively indicates the user's type such as race, ethnicity and susceptibility to communication outage, requirement for the minimum data rate to fulfill his/her needs, and the longest time to endure the degraded or loss of communication services.

To quickly restore or improve communication services, responsible stakeholders such as government or Internet service providers attempt to deploy their limited pool of resources, denoted as $S_{\text{max}}$, within a short period of time $T$ (e.g., the rule of \emph{``Golden 72 hours''}). In practice, these resources could be a swamp of satellite trucks, unmanned aerial vehicles, infrastructure repairing crews and many others \cite{duong2019learning,casoni2015integration}, and yet we do not assume a specific resource modality in this work. Without loss of generality, suppose time is slotted $t = 1,...,T$ and all the involved decision processes are made at the beginning of each time slot. Let $z_{n}(t)$ represent the binary decision variable whether to serve $n$ at time $t$ or not and $s_{n}$ indicate the amount of deployed resources. To this end, we model the \emph{exogenous resource placement} problem as follows.

\subsection{Exogenous Resource Placement}
Assume that a specific area $n$ only needs to be served (at most) once in the time window $T$ to restore or improve its local communication services. Then, we have the following inequality constraint related to the binary variable $z_{n}(t)$:
\begin{equation} \label{pl_sum_bi}
\begin{aligned}
&\sum_{t=1}^{T} z_{n}(t) \leq 1, \quad \forall n \in {\cal N}; \\
&z_{n}(t) \in \{0,1\}, \quad \forall n \in {\cal N}, 1 \leq t \leq T.
\end{aligned}
\end{equation}

To reflect the practical scenario that resources can be collected back and re-distributed (e.g., crews completing the repair of one infrastructure at one site can be dispatched to another site), we consider the resources $S_{\text{max}}$ are reusable but with a delay of $\delta$ time slots in light of the service or freeze-out time. Thus, the amount of resources available for placement at $t$ are quantitatively represented by 
\begin{equation} \label{state_var}
s(t)=s(t-1)-\sum_{n=1}^{N} z_{n}(t-1) \cdot s_{n}+\sum_{n=1}^{N} z_{n}(t-1-\delta) \cdot s_{n}, \quad 1 \leq t \leq T.
\end{equation}
Note that for any $t \leq 0$, $z_{n}(t) = 0$. Besides, at any time slot, the amount of usable resources should not exceed the amount of available resources, which is captured by the following constraints:
\begin{equation} \label{pl_res}
\sum_{n=1}^{N} z_{n}(t) \cdot s_{n} \leq s(t) \leq S_{\text{max}}, \quad 1 \leq t \leq T. 
\end{equation}

Upon the placement of resource $s_{n}$ at site $n$, users ${\cal U}_{n}$ share and access the (re-)gained network resources $c_n$ striving to meet their quality-of-service (QoS) requirements. Without loss of generality, $c_n$ here refers to the overall (re-)gained data rate that is proportional to the deployed resource $s_{n}$, i.e., $c_n = {\cal Q}(s_{n})$ in which ${\cal Q}(\cdot)$ is monotonically increasing and non-negative. Next, we model the problem of \emph{in-situ resource sharing} of $c_n$.
\subsection{In-situ Resource Sharing}
Provided with the overall resource $c_n$, the perceived data rate by any user $u_{k, n}$ is calculated as
\begin{equation} \label{rs_rate}
r_{k, n}=\frac{w_{k, n}}{\sum_{k=1}^{K_{n}} w_{k, n}} c_{n}, \quad \forall n \in {\cal N}
\end{equation}
where $w_{k, n}$ is the fraction of overall resource that is allocated to user $u_{k, n}$. This is a generic modeling, but it is applicable to many practical technologies. For instance, $w_{k, n}$ in LTE/LTE-A is mapped to the physical Resource Blocks and in TDMA to time slots. 

Recall that each user is characterized by the three-tuple attribute set $\pi$. With the achievable rate $r_{k, n}$ in Eq.(\ref{rs_rate}) and the time slot when $r_{k, n}$ is gained (i.e., as soon as $s_{n}$ is deployed), we define user's utility function $V_{k,n}$ by considering the following aspects. First, $V_{k,n}$ should be a non-negative and monotonically increasing function w.r.t. $r_{k, n}$. Nevertheless, further increasing $r_{k, n}$ beyond $\hat{r}_{k,n}$ should have decreasing marginal returns on his/her utility. To reflect such considerations, we adopt the well-known sigmoid function to define ${\cal R}\left(r_{k, n}, \hat{r}_{k,n}\right)$ as follows
\begin{equation} \label{rs_rate_util}
{\cal R}\left(r_{k, n}, \hat{r}_{k,n}\right) = \frac{1}{1+e^{ - \theta \cdot (r_{k,n}-\hat{r}_{k,n})}},
\end{equation}
where $\theta$ is a constant to control the steepness of the sigmoid function. Second, if $s_{n}$ is deployed too late to address user $u_{k, n}$'s timely service request (i.e., later than $d_{k,n}$), such resource placement is considered to have zero utility for $u_{k, n}$ regardless of $r_{k, n}$. That is to say, $V_{k,n} = 0$ if $t^{*} > d_{k,n}$ for $t^{*} = \{t|z_{n}(t) = 1, 1 \leq t \leq T \}$. Third, by considering user's type $\tau_{k,n}$, we can prioritize the service to socially under-represented groups which are comparably more vulnerable to degraded or loss of communication service due to disasters. Collectively, the utility function $V_{k,n}$ by taking the above three aspects into consideration can be defined as
\begin{equation} \label{rs_util}
\begin{aligned}
V_{k, n} &= \tau_{k, n} {\cal R}\left(r_{k, n}, \hat{r}_{k,n}\right) \mathds{1}\left(t^{*} \leq d_{k, n}\right) \\
&=\tau_{k, n} {\cal R}\left(r_{k, n}, \hat{r}_{k,n}\right) \sum_{t=1}^{\mathrm{d_{k,n}}} z_{n}(t)
\end{aligned}
\end{equation}
where ${\mathds{1}}(\cdot)$ is an indicator function whose result is 1 if its evaluated expression is true; and 0 otherwise. $V_{k, n}$ can be simplified because we observe that 
$\mathds{1}\left(t^{*} \leq d_{k, n}\right)$ $=$ $z_{n}(t) |_{t \leq d_{k, n}}$ $=$ $\sum_{t=1}^{\mathrm{d_{k,n}}} z_{n}(t)$.

\section{Problem Formulation}
In this work, our objective is to optimize the strategy for \emph{exogenous resource placement} and \emph{in-situ resource sharing} so as to maximize the utility of the users whose communication services are gravely impacted by the disaster. This would be given by the maximizer to the total utility of users in ${\cal N}$ as follows
\begin{equation}\label{opt}
\begin{aligned}
& \mathop {\text{Max} }\limits_{\langle\textbf{z}, \textbf{s}, \textbf{w}\rangle}
& & \sum_{n=1}^{N} \sum_{k=1}^{K_{n}} V_{k, n} \\
& \text{s.t.}
& &  \text{Eqs.}(\ref{pl_sum_bi})\sim(\ref{rs_rate}); \\
&&& 0 \leq s_{n} \leq S_{\text{max}}, \quad \forall n \in {\cal N}; \\
&&& w_{k, n} \in [0,1], \quad \forall k \in {\cal U}_{n}, \forall n \in {\cal N}.
\end{aligned}
\end{equation}
where $w_{k, n}$ is normalized within $[0,1]$, and the set of optimization variables is denoted as $\langle\textbf{z}, \textbf{s}, \textbf{w}\rangle$. Here, we coin the problem (\ref{opt}) as the Optimizer ${\cal O}$1. Note that the time instant $t$ does not mean ${\cal O}$1 is an online optimization problem because it considers no future utilities. Instead, ${\cal O}$1 has a static setup within a fixed time horizon $T$. 

Nevertheless, the offline nature of ${\cal O}$1 by no means eases the solution to it. It is observed that $V_{k, n}$ is not a concave function because of the multiplication of a non-strictly concave function ${\cal R}(\cdot)$ and an integer variable $\textbf{z}$, and the feasible set defined by the constraints forms a non-convex set. Overall, Optimizer ${\cal O}$1 is a non-convex mixed-integer non-linear programming (MINLP) problem, which as shown in the following theorem is NP-hard. 
\begin{theorem}\label{np-hard}
Optimizer ${\cal O}$1 is reducible to a Multiple-Choice Knapsack Problem (MCKP) \cite{moser1997algorithm}.
\end{theorem}
\begin{proof}
MCKP specifies the following problem. Suppose there are $N$ classes of items to be packed to a knapsack of finite capacity. Each item is associated with a utility and weight. The goal is to choose one item from each class and put them in the knapsack so that the overall utility is maximized while the total weight does not exceed the knapsack's capacity.

In Optimizer ${\cal O}$1, we consider each area $n \in {\cal N}$ as one class of $(T+1) \cdot S_{\text{max}} \cdot W_{n}$ items which corresponds to $z_n(t)$'s $T$ time instants (an additional time instant represents a NULL state meaning area $n$ is never served within $T$), $s_n$'s up to $S_{\text{max}}$ amount of allocated resource, and $W_{n}$ number of combinations of $w_{k,n}$ for $1 \leq k \leq K_{n}$. Optimizer ${\cal O}$1 aims to choose the optimal strategy $\langle{z^{*}}, {s^{*}}, {w^{*}}\rangle$ (i.e., ``item'') at each area (i.e., ``class'') to be placed in the knapsack so that the aggregated utility over all areas is maximized. Besides, the constraints specified by Eq.(\ref{pl_sum_bi}) and Eq.(\ref{state_var}) pose resource limits which are analogously equivalent to the capacity of the knapsack. Therefore, Optimizer ${\cal O}$1 is an MCKP instance, which is NP-hard.
\end{proof}

To solve Optimizer ${\cal O}$1, we start with transforming the utility function in Eq.(\ref{rs_util}) into the following form
\begin{equation} \label{rs_util_new}
\begin{aligned}
V_{k, n} &\mathop  = \limits^{(a)} \sum_{t=1}^{\mathrm{d_{k,n}}} \tau_{k, n} {\cal R}\left(r_{k, n}, \hat{r}_{k,n}\right) z_{n}(t) \\
&\mathop \approx \limits^{(b)} \sum_{t=1}^{\mathrm{d_{k,n}}} \frac{\tau_{k, n}}{1+e^{ - \theta \cdot \left[z_{n}(t) \cdot r_{k,n}-\hat{r}_{k,n}\right]}} \\
&\mathop = \limits^{(c)} \sum_{t=1}^{\mathrm{d_{k,n}}} \frac{\tau_{k, n} }{1+e^{ - \theta \cdot \left[z_{n}(t) \cdot w_{k,n} \cdot c_{n}-\hat{r}_{k,n}\right]}}.
\end{aligned}
\end{equation}
The reason for equality (a) is because $\tau_{k, n} {\cal R}\left(r_{k, n}, \hat{r}_{k,n}\right)$ are irrelevant to the time instant $t$. Additionally, by assuming a sufficiently large $\theta$, $V_{k, n}$ can be arbitrarily close to zero so we make an approximation by integrating $z_{n}(t)$ into the sigmoid function as in equality (b). Without loss of generality, suppose $w_{k, n}$ is normalized within $[0,1]$. Then $\sum_{k=1}^{K_{n}} w_{k, n} = 1$ and we can transform the fractional term in Eq.(\ref{rs_rate}) to a linear one as shown in equality (c).

Next, we proceed to simplify the recursive formulation of $s(t)$ in Eq.(\ref{state_var}). Specifically, by tracing time instants from $t^{'} = 1$ to $t^{'} = t$ and re-organizing Eq.(\ref{state_var}), we can have the following
\begin{equation*}
\begin{aligned}
s(t) &= s(1)-\sum_{n=1}^{N} \sum_{t^{\prime} = 1}^{t-1} \left[z_{n}\left(t^{\prime}\right)-z_{n}\left(t^{\prime}-\delta\right)\right] s_{n} \\
&=s(1)-\sum_{n=1}^{N} \sum_{t^{\prime}=t-\delta}^{t-1} z_{n}\left(t^{\prime}\right) s_{n}.
\end{aligned}
\end{equation*}
Given the boundary condition $s(1) = S_{\text{max}}$ and non-negativity of $z_{n}\left(t\right) s_{n}$, $s(t) \leq S_{\text{max}}$ always holds. Besides, by further examining $\sum_{n=1}^{N} z_{n}(t) s_{n} \leq s(t)$, we can re-write Eq.(\ref{pl_res}) into 
\begin{equation} \label{pl_res_old}
\sum_{n=1}^{N} \sum_{t^{\prime}=t-\delta}^{t} z_{n}\left(t^{\prime}\right) s_{n} \leq S_{\text{max}}, \quad 1 \leq t \leq T.
\end{equation}
Moreover, since $\sum_{k=1}^{K_{n}} w_{k, n} = 1$, we can transform the above inequality into
\begin{equation} \label{pl_res_new}
\begin{aligned}
S_{\text{max}} &\geq \sum_{n=1}^{N} \sum_{t^{\prime}=t-\delta}^{t} z_{n}\left(t^{\prime}\right) s_{n} \sum_{k=1}^{K_{n}} w_{k, n} \\
&= \sum_{n=1}^{N} \sum_{k=1}^{K_{n}} \sum_{t^{\prime}=t-\delta}^{t} z_{n}\left(t^{\prime}\right)w_{k, n} s_{n}
\end{aligned}
\end{equation}

Obviously\footnote{For the sake of brevity, we assume $s_{n} = c_{n}$ in this work.}, the transformed objective function in Eq.(\ref{rs_util_new}) and the constraint in Eq.(\ref{pl_res_new}) now have a common multivariate variable $z_{n}(t) w_{k, n} s_{n}$, which, according to the symbolic reformulation rules \cite{misener2014antigone}, can be replaced with a univariate variable $x_{k,n}(t)$. This substitution presents us the following optimization problem coined as Optimizer ${\cal O}2$ 
\begin{equation}\label{opt_new}
\begin{aligned}
& \mathop {\text{Max} }\limits_\textbf{x}
& & \sum_{n=1}^{N} \sum_{k=1}^{K_{n}} \sum_{t=1}^{\mathrm{d_{k,n}}} \frac{\tau_{k, n} }{1+e^{ - \theta \cdot \left[x_{k,n}(t)-\hat{r}_{k,n}\right]}} \\
& \text{s.t.}
& &  \sum_{n=1}^{N} \sum_{k=1}^{K_{n}} \sum_{t^{\prime}=t-\delta}^{t} x_{k,n}(t^{\prime}) \leq S_{\text{max}}, \quad 1 \leq t \leq T;\\
&&& 0 \leq x_{k,n}(t) \leq S_{\text{max}}, \quad \forall k \in {\cal U}_{n}, \forall n \in {\cal N}, 1 \leq t \leq T.
\end{aligned}
\end{equation}
With the above relaxation steps, Optimizer ${\cal O}2$ becomes a standard knapsack problem with a sigmoid objective and linear constraints w.r.t. $x_{k,n}(t)$.

\section{Heuristic Algorithm and Benchmark Scheme}
\subsection{Heuristic Algorithm}
Due to non-concavity of the sigmoid function, state-of-the-art efforts in coping with problems similar to Optimizer ${\cal O}2$ are mainly based on parametric transformation, concave envelope substitution and other relaxations \cite{srivastava2014knapsack,udell2013maximizing}. After converting the sigmoid function to an approachable form, recursive algorithms such as branch and bound, despite slow convergence speed, can be applied to determine optimal solutions. In this work, in light of the prohibitively large optimization space, we propose a near-optimal and efficient heuristic algorithm based on the spatial-temporal characteristics of the empirical dataset.

\begin{figure}[!htb]
  \begin{center}
  \includegraphics[width=1.8in]{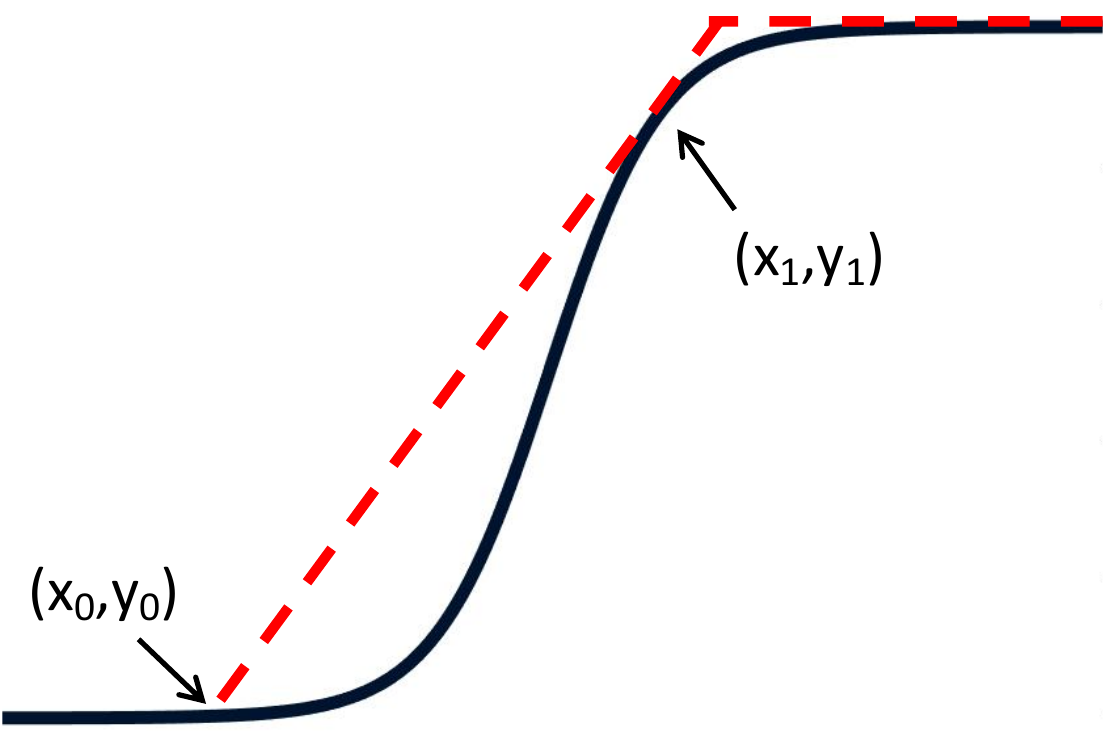}
  \end{center}
  \begin{center}
   \parbox{8cm}{\caption{Concave envelope of a sigmoid using two affine functions.}\label{approx}}
  \end{center}
\end{figure}

First, we start with replacing the sigmoid objective in Optimizer ${\cal O}2$ with its concave envelope. As shown in Fig.\ref{approx}, we can utilize two affine functions to approximate any arbitrary sigmoid; and obviously, the steeper the sigmoid (reflected by $\theta$) is, the more accurate the approximation becomes.
Specifically, $(x_0, y_0)$ can be easily determined by solving $y_0$ given $x_0 = 0$ whereas $(x_1, y_1)$ can be found by bisection. In so doing, we can re-write each sigmoid in Optimizer ${\cal O}2$'s objective function into $\min \{a_{k,n,t}x_{k,n}(t)+b_{k,n,t}, \tau_{k,n}\}$. Now, Optimizer ${\cal O}2$ becomes a maximization of the sum of piece-wise affine functions, which is equivalent to maximizing the sum of auxiliary variables $\beta_{k,n,t}$ \cite{boyd2004convex}, presented below:
\begin{equation}\label{opt_new2}
\begin{aligned}
& \mathop {\text{Max} }\limits_{\langle\textbf{x}, \boldsymbol\beta\rangle}
& & \sum_{n=1}^{N} \sum_{k=1}^{K_{n}} \sum_{t=1}^{\mathrm{d_{k,n}}} \beta_{k,n,t} \\
& \text{s.t.}
& & \beta_{k,n,t} \leq a_{k,n,t}x_{k,n}(t)+b_{k,n,t}, \\
&&& \beta_{k,n,t} \leq \tau_{k,n}, \\
&&&  \sum_{n=1}^{N} \sum_{k=1}^{K_{n}} \sum_{t^{\prime}=t-\delta}^{t} x_{k,n}(t^{\prime}) \leq S_{\text{max}}, \quad 1 \leq t \leq T;\\
&&& 0 \leq x_{k,n}(t) \leq S_{\text{max}}, \quad \forall k \in {\cal U}_{n}, \forall n \in {\cal N}, 1 \leq t \leq T.
\end{aligned}
\end{equation}
Then, addressing Optimizer ${\cal O}2$ is boiled down into solving an approachable linear programming problem in (\ref{opt_new2}). 

The solution to problem (\ref{opt_new2}) serves as an upper bound for the original Optimizer ${\cal O}1$ in (\ref{opt}). Next, by iteratively rounding the optimal solution $\mathbf{x^{*}}$ to problem (\ref{opt_new2}), we attempt to derive feasible solution that conforms to the original constraints in (\ref{opt}). Rather than adopting well-known algorithms such as (spatial) branch and bound \cite{misener2014antigone}, we develop a spatial-temporal iteration algorithm to search for feasible yet near-optimal solution. Our basic idea is as follows. First, we propose a variable $\gamma_{n,t} = \sum_{k=1}^{K_{n}} [V_{k, n} / x_{k,n}(t)]$ named as \emph{utility-resource ratio} (URR) for each area $n \in {\cal N}$ at time instant $1 \leq t \leq T$. The implication of $\gamma_{n,t}$ is to measure how effective allocating resource can produce utility gain. Then, motivated by the principle that each area is served at most once before time $T$, the \emph{temporal policy} is set to serve an area at the time instant $t^{*}$ when its URR is maximum, that is $t^{*} =  \arg \max \limits_t \gamma_{n,t}, \, \forall n \in {\cal N}$. If there is a tie, we resolve it by following the \emph{spatial policy} that asserts a time instant $\hat{t}$ when the area's URR, comparing to other areas', ranks the highest among the tied time instants. Mathematically, $\hat{t} = \arg \max \limits_{t^{*}} \text{Rank}(n, \gamma_{{\cal N},t^{*}})$ where the function $\text{Rank}()$ gives the index of $n$ in a non-decreasingly sorted vector $\gamma_{{\cal N},t^{*}}$. If tie persists, we select the earliest time instant to serve the area, which is termed as \emph{tie-breaker policy}. For presentation clarity, we summarize the steps in the following algorithm.
\begin{algorithm}
\caption{Spatial-Temporal Rounding Algorithm}\label{algo}
\begin{algorithmic}[1]
    \REQUIRE Resource $S_\text{max}$, user's attribute set $\pi = \langle\tau, \hat{\mathrm{r}}, \mathrm{d}\rangle$, areas of interests $\cal N$ and $T$.
    \ENSURE $\langle\textbf{z}, \textbf{s}, \textbf{w}\rangle$
\STATE Determine $\textbf{x}$ by solving Problem (\ref{opt_new2}).
\STATE Set $\textbf{z} = \textbf{0}$ (or $=\textbf{1}$) by examining if $\textbf{x} = \textbf{0}$ (or $\neq \textbf{0}$).
\IF { $\exists n \in {\cal N},$ $\sum_{t=1}^{T} z_{n}(t) > 1$}
\IF {$\exists n \in {\cal N},$ it has different non-zero URRs across $[1,T]$}
\STATE Apply the \emph{temporal policy}, then go to Step 1;
\ELSE
\STATE Apply the \emph{spatial policy} and the \emph{tie-breaker policy} if needed, then go to Step 1;
\ENDIF
\ELSE
\STATE Calculate $\langle\textbf{z}, \textbf{s}, \textbf{w}\rangle$ from $\textbf{x}$.
\ENDIF
\end{algorithmic}
\end{algorithm}

\subsection{Benchmark Scheme}
The state-of-the-art modeling on post-disaster communication provisioning mainly aims for serving the maximum number of users (i.e., via admission control) \cite{beard2001prioritized}, maximizing the aggregated throughput at disaster sites \cite{duong2019learning}, prioritizing critical service requests \cite{lieser2019understanding}, and etc. To examine how significant our design can promote services for under-represented users, we present a well-accepted model from related works as a comparing benchmark.

Since existing works may consider different resource constraints than ours, to make a fair comparison, we adjust the benchmark model to our context while preserving its objective. Specifically, the adopted benchmark model is boiled down into the following optimization problem, coined as ${\cal B}{\cal O}$:
\begin{equation}\label{opt_benchmark}
\begin{aligned}
& \mathop {\text{Max} }\limits_{\langle\textbf{z}, \textbf{s}, \textbf{w}\rangle}
& & \sum_{n=1}^{N} \sum_{k=1}^{K_{n}} \sum_{t=1}^{\mathrm{d_{k,n}}} z_n(t) \\
& \text{s.t.}
& & \text{Eq.}(\ref{pl_sum_bi}), \\
&&& \sum_{n=1}^{N} \sum_{k=1}^{K_n} \sum_{t^{\prime}=t-\delta}^{t} z_{n}\left(t^{\prime}\right) s_{k, n} \leq S_{\text{max}}, \quad 1 \leq t \leq T,\\
&&& \sum_{t=1}^{\mathrm{d_{k,n}}} z_n(t) s_{k, n} \geq \hat{r}_{k,n} \cdot \phi, \quad \forall k \in {\cal U}_{n}, \forall n \in {\cal N}. \\
\end{aligned}
\end{equation}
The objective tells that ${\cal B}{\cal O}$ aims to serve as many users as possible before their respective deadline. The first two constraints are inherited from our problem setting, whereas the third constraint reflects the concept of \emph{service differentiation}, e.g., partial service for non-critical requests and full service for critical ones (controlled by $\phi$) \cite{savas2014network}. Note that for the sake of easiness in later calculation, we replace $w_{k, n} s_{n}$ with an auxiliary variable $s_{k, n}$ and $s_{n}$ with $\sum_{k=1}^{K_N} s_{k, n}$.

The major obstacle to solving ${\cal B}{\cal O}$ lies in the integer variable $z_n(t)$ and the bivariate variable $z_n(t)s_{k, n}$ (i.e., the quadratic term). On the one hand, $s_{k, n}$ is loosely coupled in the objective function so replacing $z_n(t)s_{k, n}$ with a univariate variable, similar to how we handle Optimizer ${\cal O}$2 in (\ref{opt_new}), does not work well. On the other hand, it is not difficult to prove that despite relaxing $z_n(t)$, solving ${\cal B}{\cal O}$ as quadratic programming is infeasible (i.e., NP-hard to be precise) since the objective's coefficient matrix which is obtained after applying partial Lagrangian transformation to the second inequality is indefinite. In light of it, our proposed solution to ${\cal B}{\cal O}$ is via the classic BnB algorithm w.r.t. $z_n(t)$. While ensuring the third inequality in ${\cal B}{\cal O}$, we will evenly distribute the remaining resource among $k \in {\cal U}_{n}$ . 
 
\section{Performance Evaluation}
\subsection{Dataset Construction}
We used data collected from household surveys that were distributed to residents in Harris County, Texas after Hurricane Harvey \cite{esmalian2020empirical}. Harvey caused catastrophic flooding and landfall in Harris County and resulted in severe disruptions in communication services. The designed survey includes, but is not limited to, questions like (1) what is the hardship your household experiences for disruptions in communication services after Hurricane Harvey (represented as \emph{hardship} throughout the paper); (2) how many days your household is capable of tolerating the interruptions on communication services (represented as \emph{tolerance} throughout the paper); (3) what are the primary needs of your household related to communication services; and many others (represented as \emph{service demand}, i.e., QoS, throughout the paper); (4) what is your household's perception on the importance of Internet or other forms of communication services after a disaster (represented as \emph{perceptions on Internet importance} throughout the paper). Along with these communication-related questions, we also acquired responders' demographic information such as their race, ethnicity, income and education level.
\begin{figure}[!htb]
  \begin{center}
  \includegraphics[width=3.1in]{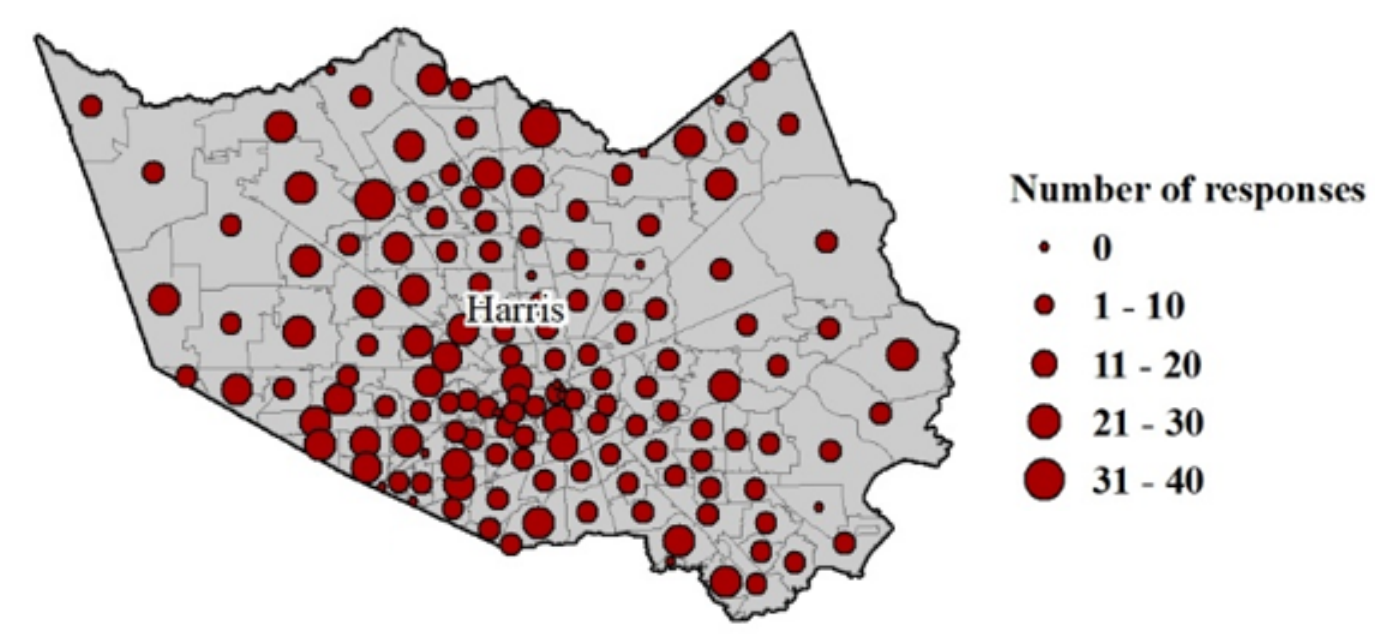}
  \end{center}
  \begin{center}
   \parbox{8cm}{\caption{Geographical distribution of survey responses.}\label{survey}}
  \end{center}
\end{figure}

In total, we obtained 872 valid records from the survey, which are geographically distributed in Harris County as shown in Fig.\ref{survey} \cite{esmalian2021susceptibility}. Among these collected data, 412 records are trimmed because these households either evacuated before Hurricane Harvey or experienced no communication service outage at their specific areas. The remaining 460 records are distributed in 125 zipcode areas (from 77001 to 77598). By combining geographically adjacent areas, we coded the dataset into 60 discrete areas of interest.

\subsection{Simulation Setup}
To turn the constructed dataset into simulation input, we quantify several descriptive attributes of the dataset as shown in Table \ref{code_value}. 
\begin{table}[h]\small
\caption{Quantitative measure of descriptive attributes}
\label{code_value}
\centering
\begin{tabularx}{0.49\textwidth}{X||l||l}
\hline
\textbf{Attribute} & \textbf{Description} & \textbf{Coded Value}\\
\hline
Income ($\tau$) & Above \$100,000 & 1 \\
 & \$49,999 - \$99,999 & 3 \\
 & Less than \$49,999 & 5 \\
\hline
Race/Ethnicity ($\tau$) & White & 1 \\
 & Non-White & 3 \\
\hline
Education ($\tau$) & Graduate school & 1 \\
 & Bachelor & 2 \\
  & Some college & 4 \\
  & High school & 5 \\
  & Less than high school & 7 \\
\hline
Service Demand ($\hat{r}$, in Mbps) & Communicate with family & 1 \\
 & Use social media & 10 \\
  & Remote work or education & 500 \\
  & Streaming entertainment & 1,000 \\
\hline
Hardship & A little & 1 \\
 & A moderate amount & 2 \\
  & A lot & 3 \\
  & A great deal & 4 \\
\hline
Perceptions on Internet Importance & Not important & 1 \\
 & Slightly important & 2 \\
  & Moderately important & 3 \\
  & Very important & 4 \\
  & Extremely important & 5 \\
\hline
\end{tabularx}
\end{table}
Every household's tolerance level to communication service disruption is found to be within 14 days for all survey responses, so we set the responding time window $T = 15$ and each time instant represents one day. Moreover, the exogenous resource parameters $S_{\text{max}}$ and $\delta$ are considered as system variables, and $S_{\text{max}}$ has a unit of Gbps which in practice can be deemed as the total amount of available bandwidth to be provisioned. In the benchmark scheme, we set $\phi = 1$ for service demand of communication with family, which is deemed more critical, while $\phi = 0$ for other service demands. 

The solution to related optimization problems is carried in MATLAB v.R2020a on a iMAC host machine with 4.2GHz quad-core and 16GB memory. Specifically, we apply the interior-point algorithm (embedded in the \texttt{linprog} toolbox) to solve Problem (\ref{opt_new2}) and implement the Algorithm \ref{algo} to approximate the solution to Problem (\ref{opt}).

\subsection{Result Analysis}
\begin{figure*}[!t]
\begin{subfigure}[t]{0.245\textwidth}
  \includegraphics[width=\linewidth]{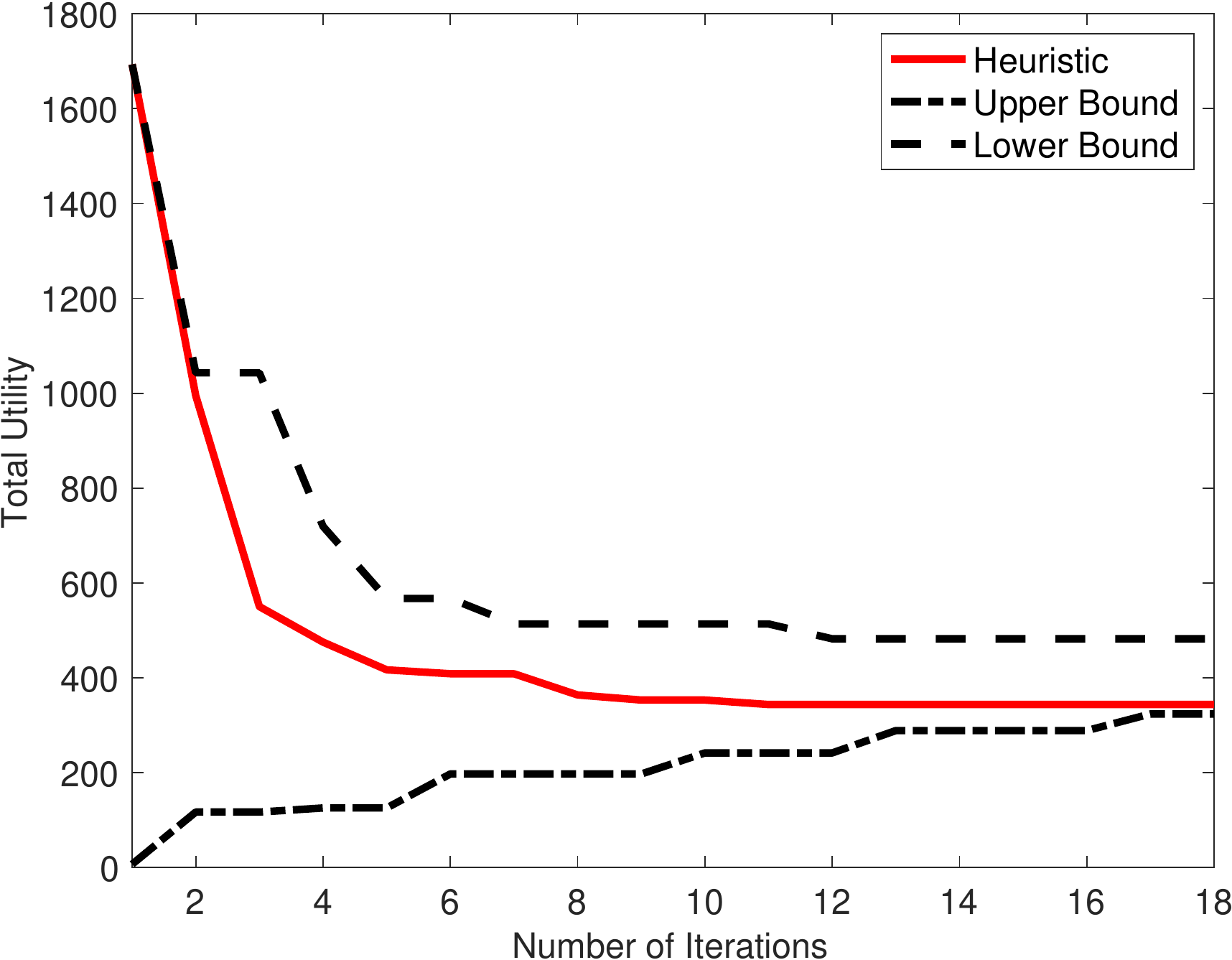}
  \caption{\small Convergence and optimality of the heuristic algorithm.} \label{conv}
\end{subfigure}
\begin{subfigure}[t]{0.242\textwidth}
  \includegraphics[width=\linewidth]{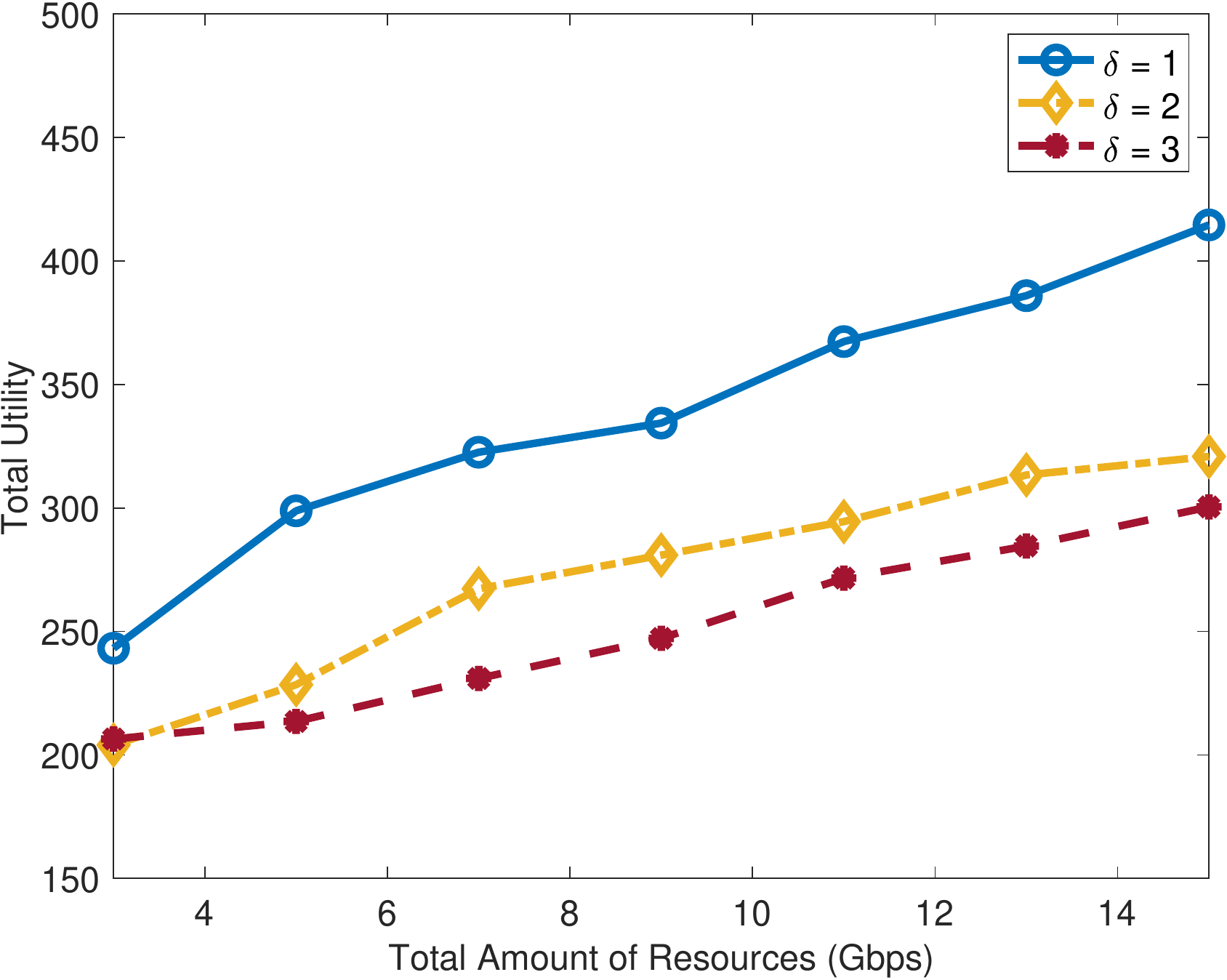}
  \caption{\small Total user utility across time and areas.} \label{tot_util}
\end{subfigure}
\begin{subfigure}[t]{0.238\textwidth}
  \includegraphics[width=\linewidth]{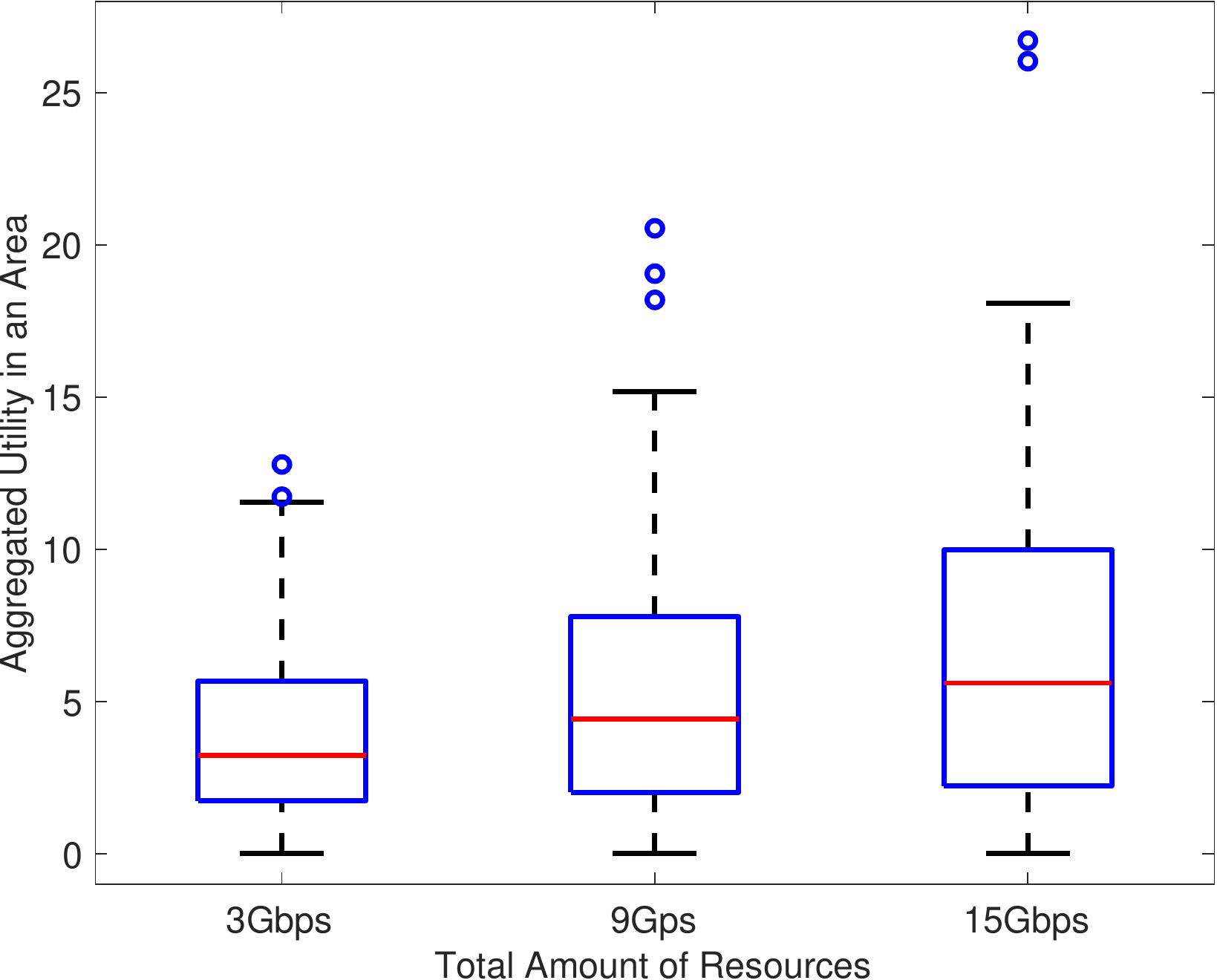}
  \caption{\small Aggregated user utility in each area of interest.} \label{loc_util}
\end{subfigure}
\begin{subfigure}[t]{0.25\textwidth}
  \includegraphics[width=\linewidth]{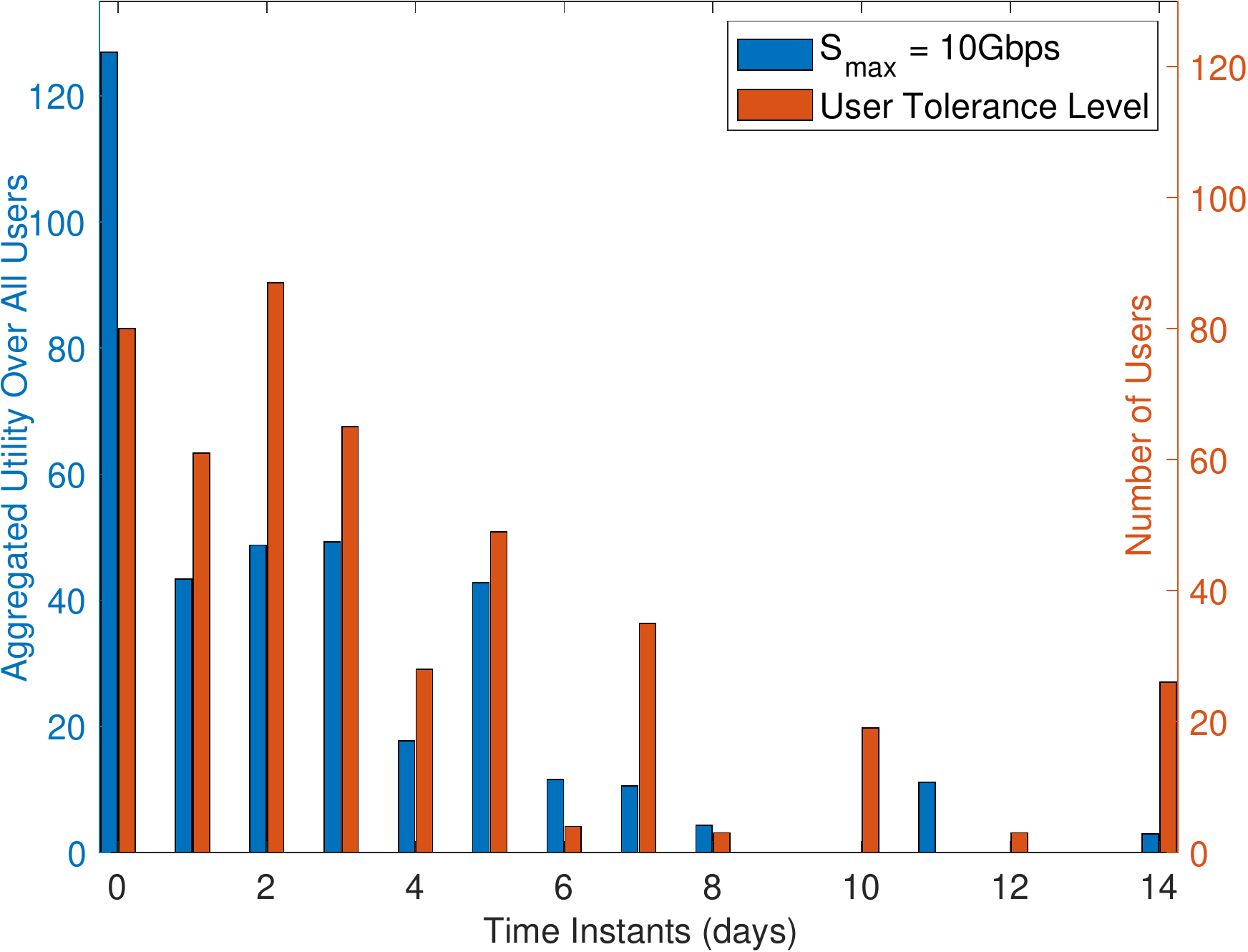}
  \caption{\small Aggregated user utility in each time instant.} \label{time_util}
\end{subfigure}
   \caption{Convergence, optimality, and spatial \& temporal performance of our design w.r.t. exogenous resource.} \label{util_eval}
\end{figure*}

\begin{figure*}[!t]
\centering
\begin{subfigure}[t]{0.275\textwidth}
  \includegraphics[width=\linewidth]{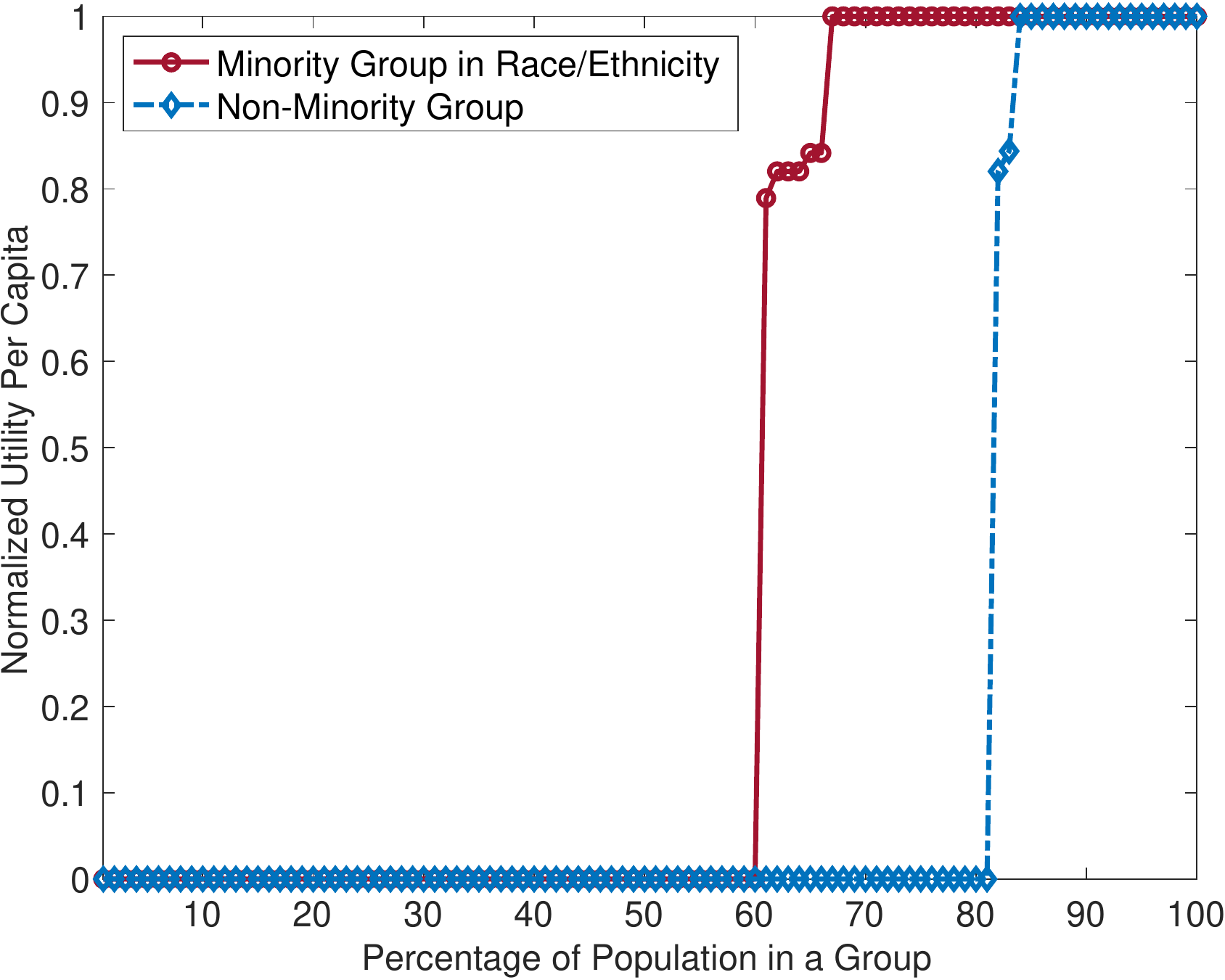}
  \caption{\small $\tau$ represents race/ethnicity.} \label{util_race}
\end{subfigure}
\begin{subfigure}[t]{0.275\textwidth}
  \includegraphics[width=\linewidth]{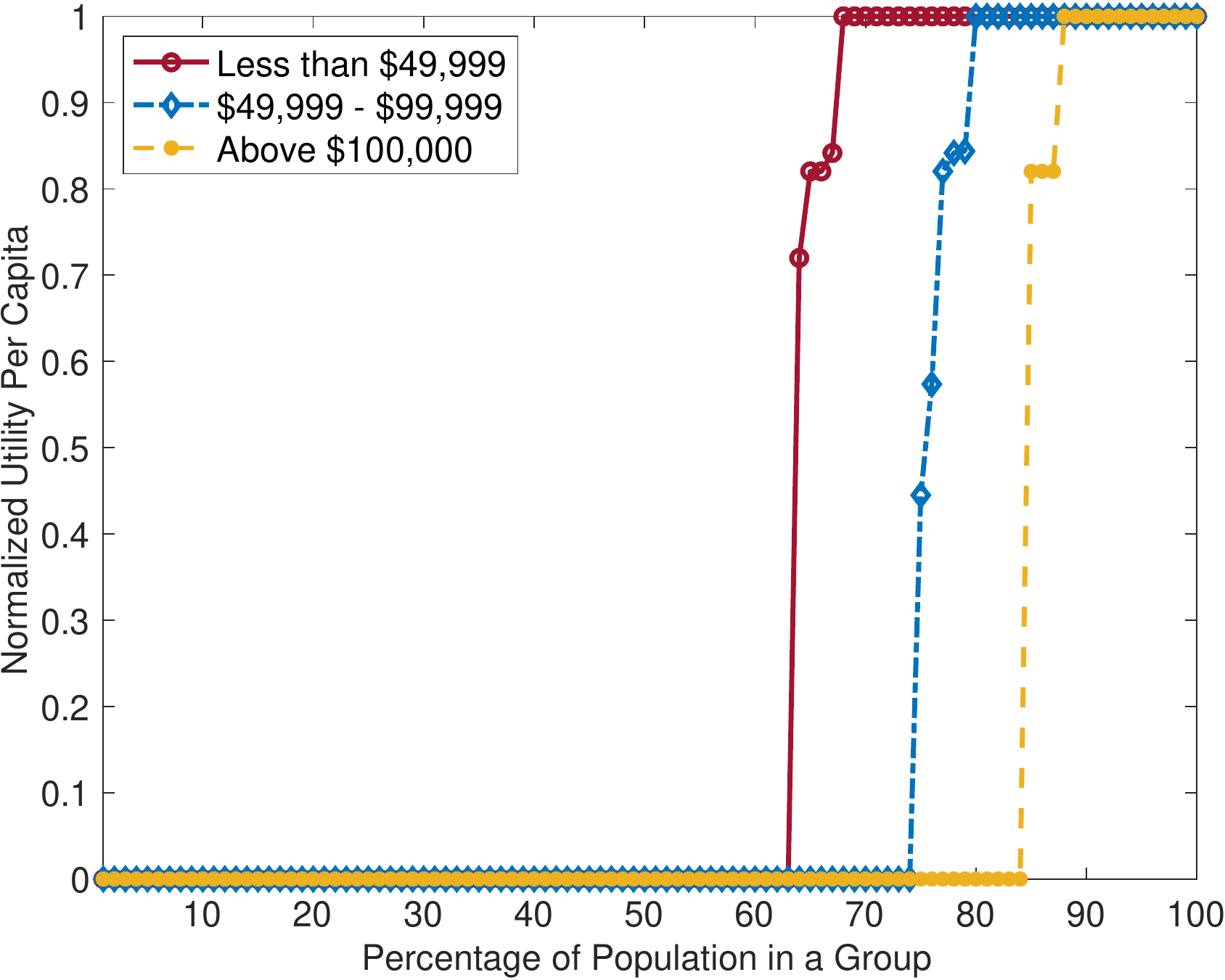}
  \caption{\small $\tau$ represents income level.} \label{util_income}
\end{subfigure}
\begin{subfigure}[t]{0.275\textwidth}
  \includegraphics[width=\linewidth]{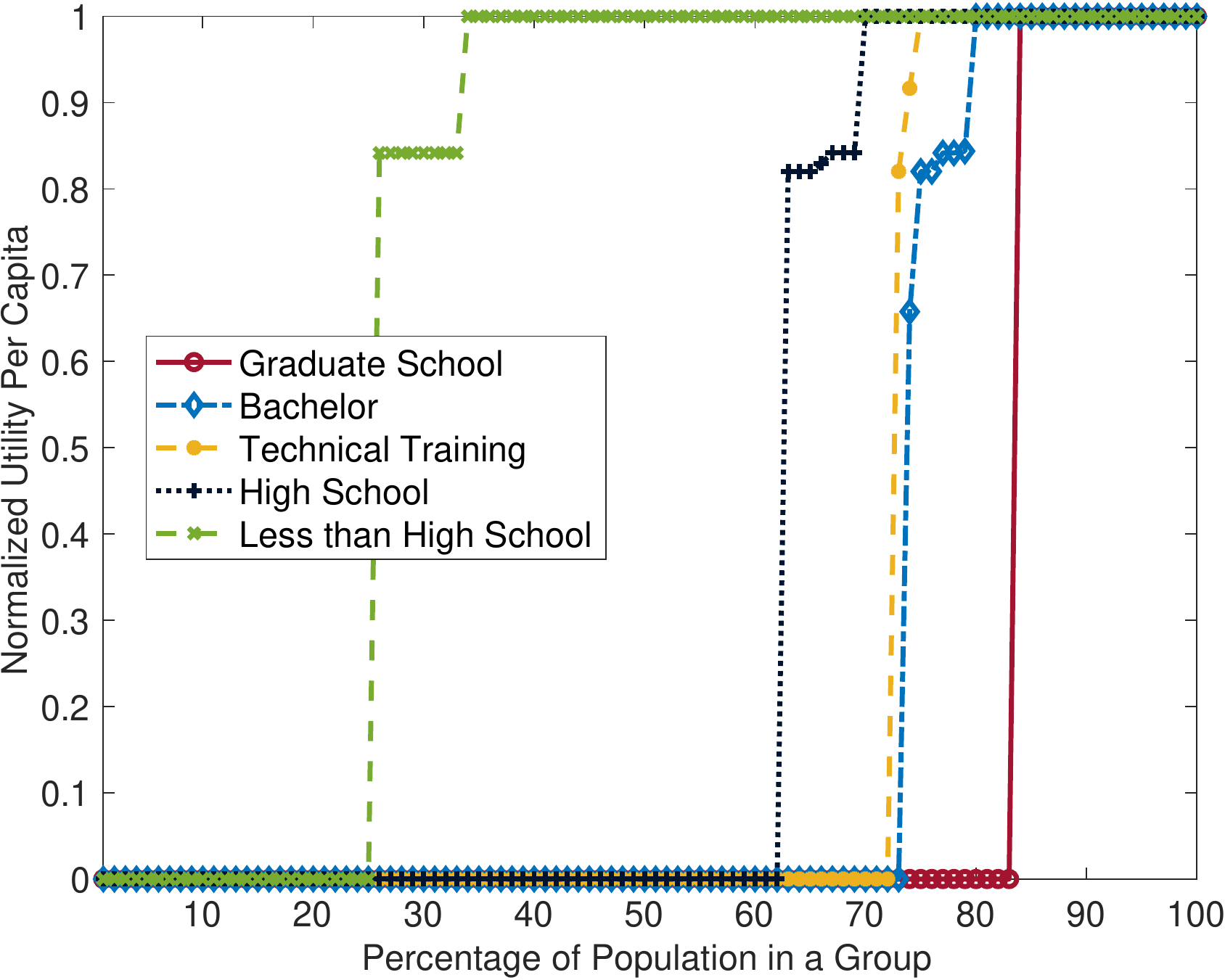}
  \caption{\small $\tau$ represents education level.} \label{util_edu}
\end{subfigure}
   \caption{User's utility under various societal factors $\tau$.} \label{usr_eval}
\end{figure*}

\subsubsection{Convergence and Optimality Analysis}
First of all, we attempt to evaluate the convergence speed of our developed heuristic algorithm and examine how close the derived solution is to the optimal one. The baseline algorithm we adopted is the well-known spatial branch and bound (S-BnB) algorithm \cite{misener2014antigone} which by theory promises to find the optimal solution to a mix-integer (non-) linear programming problem. Specifically, we apply the S-BnB algorithm to tackle Problem (\ref{opt_new2}) and note that the obtained solution only serves as an upper bound to the original problem in (\ref{opt}) because of the objective's relaxation based on the concave envelope in Fig.\ref{approx}. 

The simulation sets $S_{\text{max}} = $10 Gbps, $\delta =$ 1 day and $\tau$ representing race/ethnicity. As shown in Fig.\ref{conv}, our heuristic algorithm converges after 12 rounds of iterations (less than 20 seconds in MATLAB) at which S-BnB (when we set its convergence criteria $\alpha$, defined as (upper bound - lower bound)/upper bound, to be 15\%) has not completed its running yet. When S-BnB converges after over 27 iterations, the result shows that our heuristic algorithm falls between its upper and lower bound, which means that our solution differs no larger than 15\% from the optimal solution. 

\subsubsection{How exogenous resource impacts users' utility}
Next, we will examine how the characteristics (i.e., amount and freeze-out time) of exogenous resource could affect the communication service restoration. In this part of simulation, we set $\tau$ to represent race/ethic and let $S_{\text{max}}$ and $\delta$ be control variables. Intuitively, the larger the amount of resource or the smaller the freeze-out time, the larger the total utility that will be obtained. $S_{\text{max}}$ and $\delta$ can be considered to have the homogeneous impact on users' utility because the smaller resource re-usable freeze-out time implies that there are more available resources at any time instant. This is reflected by our simulation results in Fig.\ref{tot_util}. In fact, users are under-served by the amount of resources (2-15 Gbps) that are examined in Fig.\ref{tot_util} because increasing $S_{\text{max}}$ always leads to the linear increase in users' utility. While when all users' services are satisfied, further increasing $S_{\text{max}}$ will ultimately cause the decreasing marginal gains on users' utility. Nonetheless, we believe our simulation in Fig.\ref{tot_util} reflects the practical post-disaster restoration scenario where resources are scarce and limited.

The above argument can be further supported by our findings as shown in Fig.\ref{loc_util}, in which we present how well different areas of interest are served by the exogenous resource. Although larger $S_{\text{max}}$ leads to higher per-area utility, some areas always receive zero utility meaning no resources are ever deployed there. By looking into these under-served areas and cross-checking the dataset, we notice that these areas are wealthy suburbs near the downtown of Houston (e.g., zipcode 77007 and 77008). Residents in these areas tend to have much higher service demand (e.g., most of the survey responders' said they need streaming entertainment) than people in other areas. Therefore, when the amount of exogenous resource is limited, the solution to Problem (\ref{opt_new2}) inclines to first allocate resources to those areas/residents whose service demands are easily (i.e., low QoS requirement) and in dire need to be addressed. 

In prior analysis, we present how resource allocation differs in space, and meanwhile, we are interested to see how resources are distributed in time. As shown in Fig.\ref{time_util}, users' tolerance level to communication disruptions is plotted in orange against the right y-axis while users' obtained utility in our solution is plotted in blue against the left y-axis. On the one hand, we observe that most users' cannot tolerate communication disruptions longer than 5 days excepts for a few outliers who can tolerate up to 14 days. Based on our analysis on the dataset, we posit that these outlier users either are not serious in answering this question (only give an estimated number) or are genuinely technology independent in their life for reasons like old age. On the other hand, the derived resource allocation strategy well coincides with the distribution of users' tolerance level, that is users' aggregated utility is heavily skewed towards the immediate aftermath of the disaster.

\subsubsection{How our design prioritizes services for socially under-represented users}
Here, we aim to promote social equality in service provisioning so we are interested to examine if communication services for minority groups could be prioritized. In the simulation, we separately consider three societal factors that define minority groups, and collectively they are race/ethnicity, income, and education as shown in Table \ref{code_value}. By design, we prioritize socially under-represented users' utility via assigning them with a larger weighting factor $\tau$. This simulation is conducted under the setting of $S_{\text{max}} = $ 10 Gbps and $\delta =$ 1 day. The results for the per-capita user utility w.r.t. different societal factors are presented in Fig.\ref{usr_eval}. Here, user utility is normalized against the maximum value of $V_{k,n}$ in Eq.(\ref{rs_util}), and specifically, when the normalized utility is above 0.5 (i.e., the inflection point of a sigmoid function), it represents user's communication service is satisfied (i.e., $r \geq \hat{r}$ and $t \leq d$). To use the case of $\tau$ as race/ethnicity as an example, our design fulfills the communication demands for around 40\% of the minority group users (e.g., Hispanic or Latino, Black, American Indian, Native Hawaiian and Pacific Islanders) and around 20\% of the non-minority group users (e.g., white), which is an evident and significant bias (or prioritization) to the minority groups. Similar trends can be observed in Fig.\ref{util_income} and Fig.\ref{util_edu}, in which $\tau$ represents income and education level, respectively. 

It is worthwhile to highlight that the most notable service prioritization occurs for users with a low education level that is \emph{less than high school}. To explain this and some other similar observations in Fig.\ref{usr_eval}, we present a correlation analysis based on the well-known Spearman's rank correlation coefficient \cite{myers2004s} and Pearson correlation coefficient \cite{benesty2009pearson}, as shown in Table \ref{spearman_correlation} and \ref{pearson_correlation}, respectively. Note that for a p-value that is less than 0.05, it indicates the evaluated pairs have a strong correlation. This is reflected by the asterisk marks in Table \ref{spearman_correlation} and \ref{pearson_correlation}. It can be observed that the experienced hardship has a strong and positive correlation with societal factors, implying that marginalized users in race, education and income are more likely to experience difficulties in communication services than their non-marginalized peers. Moreover, people of low income and low education are evidently more incline to tolerate less days of communication service outage than their high-income or high-education peers. Regarding the service demand, although it correlates negatively to the societal factors, there is no statistical significance. We have to concede that this may not reflect the truth because of the non-ideal design of our survey questions. Specifically, the question regarding service demand is designed as a multiple-choice question and it does not rank a responder's choices from high to low priority, which unfortunately creates uncertainty in our analysis. 
\begin{table}[h]\small
\caption{Spearman-based Correlation Analysis}
\label{spearman_correlation}
\centering
\begin{tabular}{l|l|c|c|c}
\hline
                                                                             &                                                               & \textbf{Income}                                          & \textbf{Education}                                       & \textbf{Race}                                 \\ \hline
Hardship                                                                     & \begin{tabular}[c]{@{}l@{}}Coefficient\\ p-value\end{tabular} & \begin{tabular}[c]{@{}c@{}}0.175*\\ 0.0002\end{tabular}  & \begin{tabular}[c]{@{}c@{}}0.111*\\ 0.0171\end{tabular}  & \begin{tabular}[c]{@{}c@{}}0.127*\\ 0.0061\end{tabular} \\ \hline
\begin{tabular}[c]{@{}l@{}}Perception \\ on Internet \\ Importance\end{tabular} & \begin{tabular}[c]{@{}l@{}}Coefficient\\ p-value\end{tabular} & \begin{tabular}[c]{@{}c@{}}-0.004\\ 0.9363\end{tabular}  & \begin{tabular}[c]{@{}c@{}}-0.057\\ 0.2260\end{tabular}  & \begin{tabular}[c]{@{}c@{}}0.027\\ 0.5620\end{tabular}  \\ \hline
Tolerance                                                                    & \begin{tabular}[c]{@{}l@{}}Coefficient\\ p-value\end{tabular} & \begin{tabular}[c]{@{}c@{}}-0.107*\\ 0.0217\end{tabular} & \begin{tabular}[c]{@{}c@{}}-0.103*\\ 0.0273\end{tabular} & \begin{tabular}[c]{@{}c@{}}-0.081\\ 0.0829\end{tabular} \\ \hline
\begin{tabular}[c]{@{}l@{}}Service \\ Demand \end{tabular}      &                       \begin{tabular}[c]{@{}l@{}}Coefficient\\ p-value\end{tabular} & \begin{tabular}[c]{@{}c@{}}-0.086\\ 0.0665\end{tabular}  & \begin{tabular}[c]{@{}c@{}}-0.019\\ 0.6806\end{tabular}  & \begin{tabular}[c]{@{}c@{}}-0.028\\ 0.5563\end{tabular} \\ \hline
\end{tabular}
\end{table}

\begin{table}[h]\small
\caption{Pearson-based Correlation Analysis}
\label{pearson_correlation}
\centering
\begin{tabular}{l|l|c|c|c}
\hline
                                                                             &                                                               & \textbf{Income}                                          & \textbf{Education}                                       & \textbf{Race}                                 \\ \hline
Hardship                                                                     & \begin{tabular}[c]{@{}l@{}}Coefficient\\ p-value\end{tabular} & \begin{tabular}[c]{@{}c@{}}0.169*\\ 0.0003\end{tabular}  & \begin{tabular}[c]{@{}c@{}}0.120*\\ 0.0100\end{tabular}  & \begin{tabular}[c]{@{}c@{}}0.156*\\ 0.0008\end{tabular} \\ \hline
\begin{tabular}[c]{@{}l@{}}Perception \\ on Internet \\ Importance\end{tabular} & \begin{tabular}[c]{@{}l@{}}Coefficient\\ p-value\end{tabular} & \begin{tabular}[c]{@{}c@{}}-0.008\\ 0.8713\end{tabular}  & \begin{tabular}[c]{@{}c@{}}-0.0556\\ 0.2346\end{tabular}  & \begin{tabular}[c]{@{}c@{}}0.0200\\ 0.6695\end{tabular}  \\ \hline
Tolerance                                                                    & \begin{tabular}[c]{@{}l@{}}Coefficient\\ p-value\end{tabular} & \begin{tabular}[c]{@{}c@{}}-0.088\\ 0.0594\end{tabular} & \begin{tabular}[c]{@{}c@{}}-0.084\\ 0.0718\end{tabular} & \begin{tabular}[c]{@{}c@{}}-0.054\\ 0.2515\end{tabular} \\ \hline
\begin{tabular}[c]{@{}l@{}}Service \\ Demand \end{tabular}      &                       \begin{tabular}[c]{@{}l@{}}Coefficient\\ p-value\end{tabular} & \begin{tabular}[c]{@{}c@{}}-0.079\\ 0.0892\end{tabular}  & \begin{tabular}[c]{@{}c@{}}-0.027\\ 0.5638\end{tabular}  & \begin{tabular}[c]{@{}c@{}}-0.028\\ 0.5426\end{tabular} \\ \hline
\end{tabular}
\end{table}

Given the above correlation analysis, we can provide more insights on our results shown in Fig.\ref{usr_eval}. First, users with low education level experience higher hardship and have lower tolerance towards communication service disruptions. Second, by examining the collected dataset, they are not heavily dependent on communication technologies for bandwidth-demanding activities such as work, education, or entertainment. Therefore, the derived solution from the optimization problem tends to prioritize resources for these low-education groups, and this leads to high utilities for them as shown in Fig.\ref{util_edu}. 
\begin{figure}[!htb]
  \begin{center}
  \includegraphics[width=2.3in]{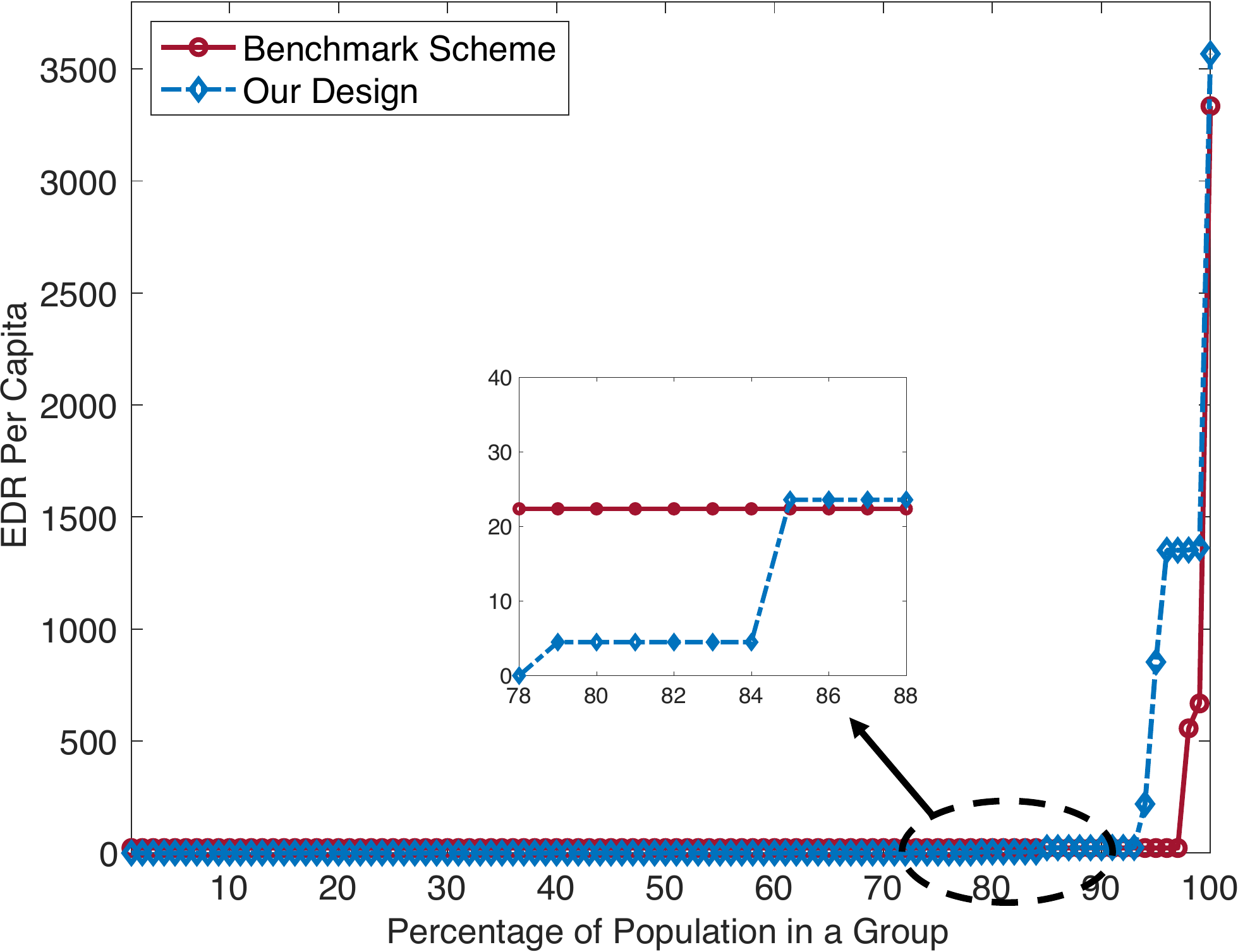}
  \end{center}
  \begin{center}
   \parbox{8cm}{\caption{Comparison of Per-capital EDR (Mbps).}\label{comp_tot}}
  \end{center}
\end{figure}

\subsubsection{How effective our design is when compared with the benchmark scheme} To evaluate this spec, we first need to have a consistent evaluation metric for comparison. Instead of adopting sigmoid as the utility function, we define \emph{effective data rate} (EDR) as following
\begin{equation} \label{edr}
\text{EDR}_{k,n} = \sum_{t=1}^{\mathrm{d_{k,n}}} z_n(t) s_{k,n}, \quad \forall k \in {\cal U}_{n}, \forall n \in {\cal N}.
\end{equation}
In this part of evaluation, we let $S_{\text{max}}=$ 10 Gbps and $\delta=$ 3 days. First, when we examine the overall resource allocation among all users, Fig.\ref{comp_tot} shows that the benchmark scheme ensures that every user is allocated with at least 1 Mbps data rate (i.e., for communications with family) but only a limited number of users (< \%5) receive above 500 Mbps data rate (i.e., for remote work, education and entertainment). On the contrary, our design offers more users (around 3-5\% more) with guaranteed services. Nevertheless, the admission rate (i.e., objective in ${\cal B}{\cal O}$) for the critical service (i.e., communications with family) in the benchmark scheme is 100\%, which could be advantageous in critical scenarios (e.g., first few hours in the aftermath of disasters).

\begin{figure*}[!t]
\centering
\begin{subfigure}[t]{0.28\textwidth}
  \includegraphics[width=\linewidth]{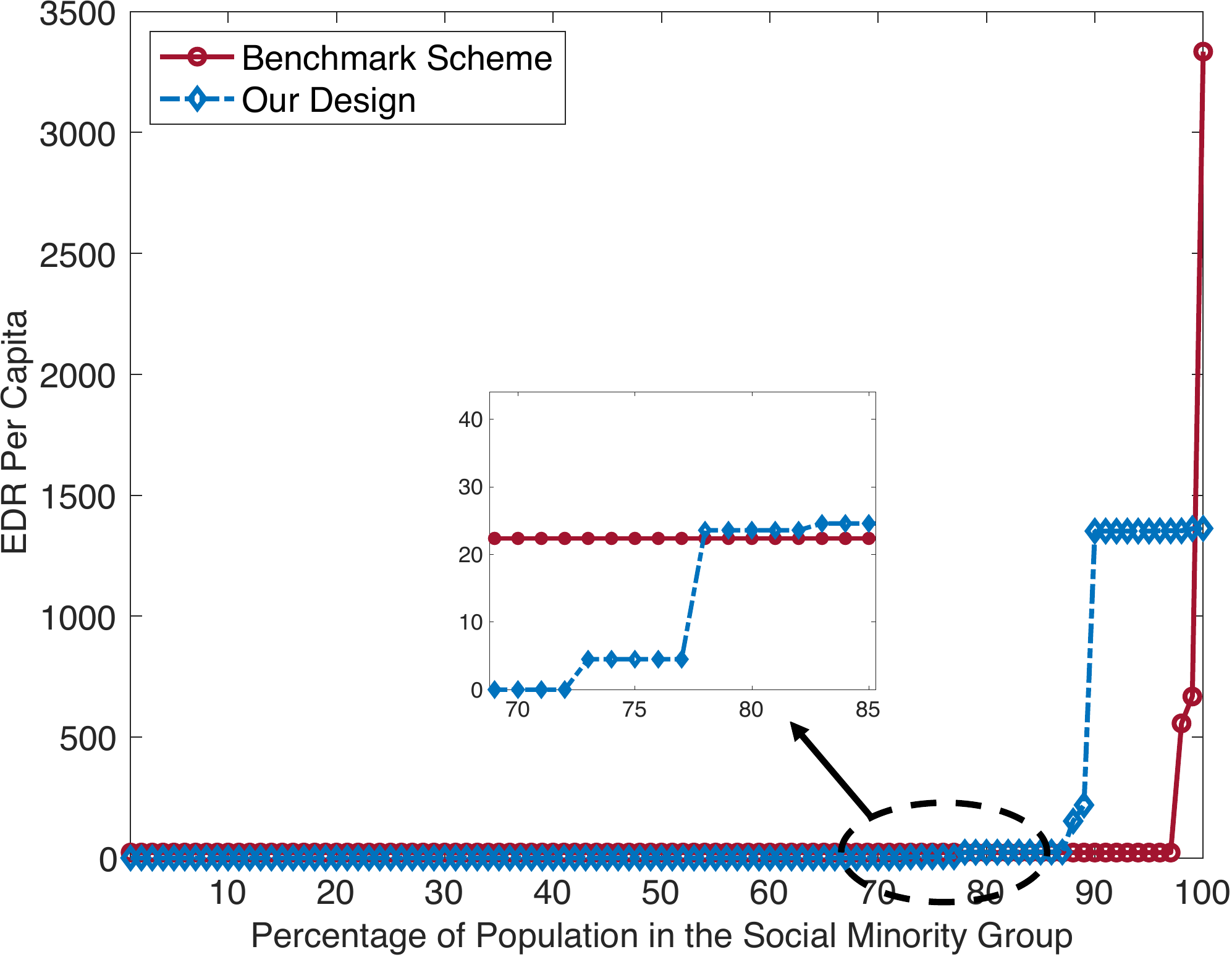}
  \caption{\small EDR in the social minority Group.} \label{comp_race}
\end{subfigure}
\begin{subfigure}[t]{0.28\textwidth}
  \includegraphics[width=\linewidth]{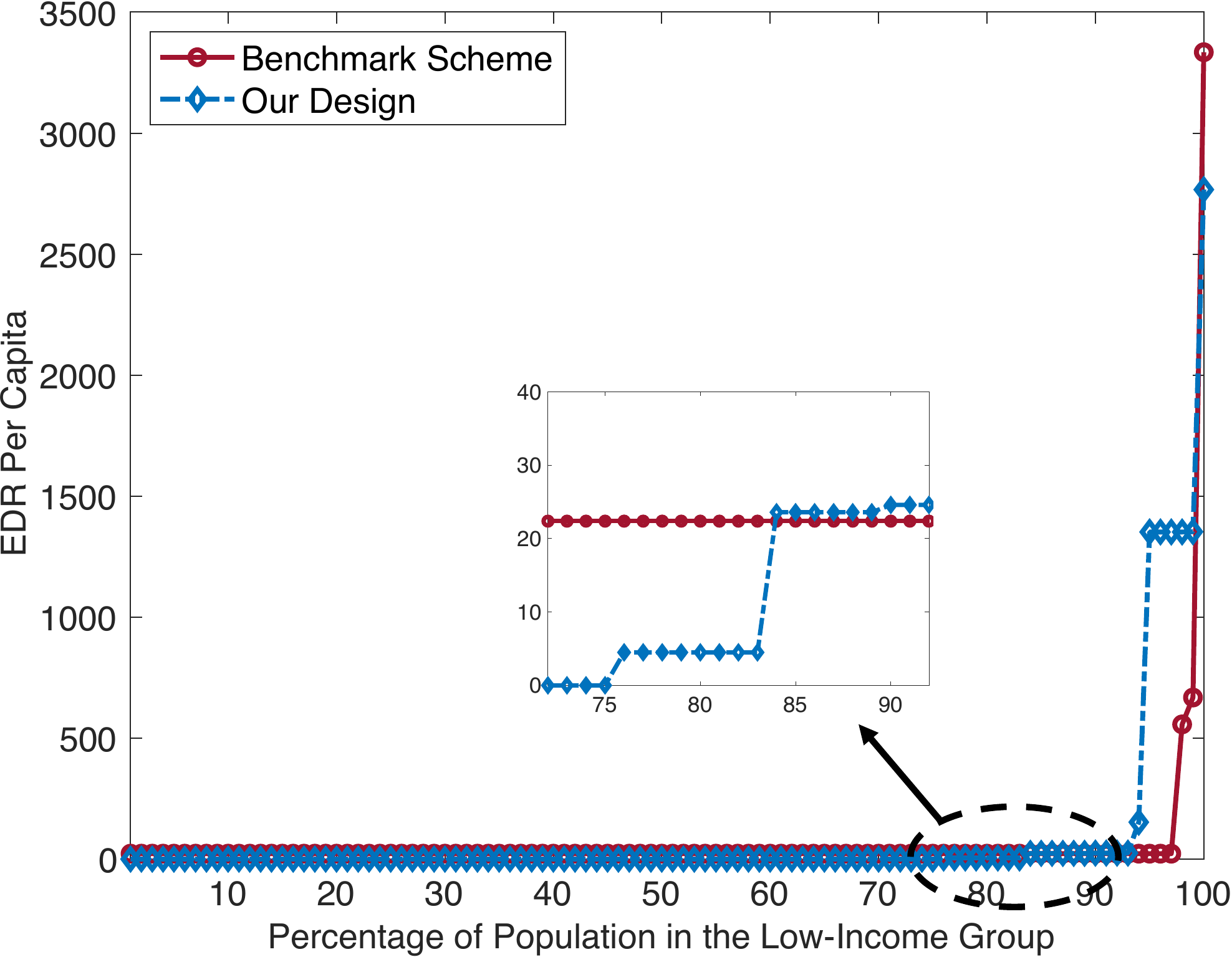}
  \caption{\small EDR in the low-income group.} \label{comp_income}
\end{subfigure}
\begin{subfigure}[t]{0.28\textwidth}
  \includegraphics[width=\linewidth]{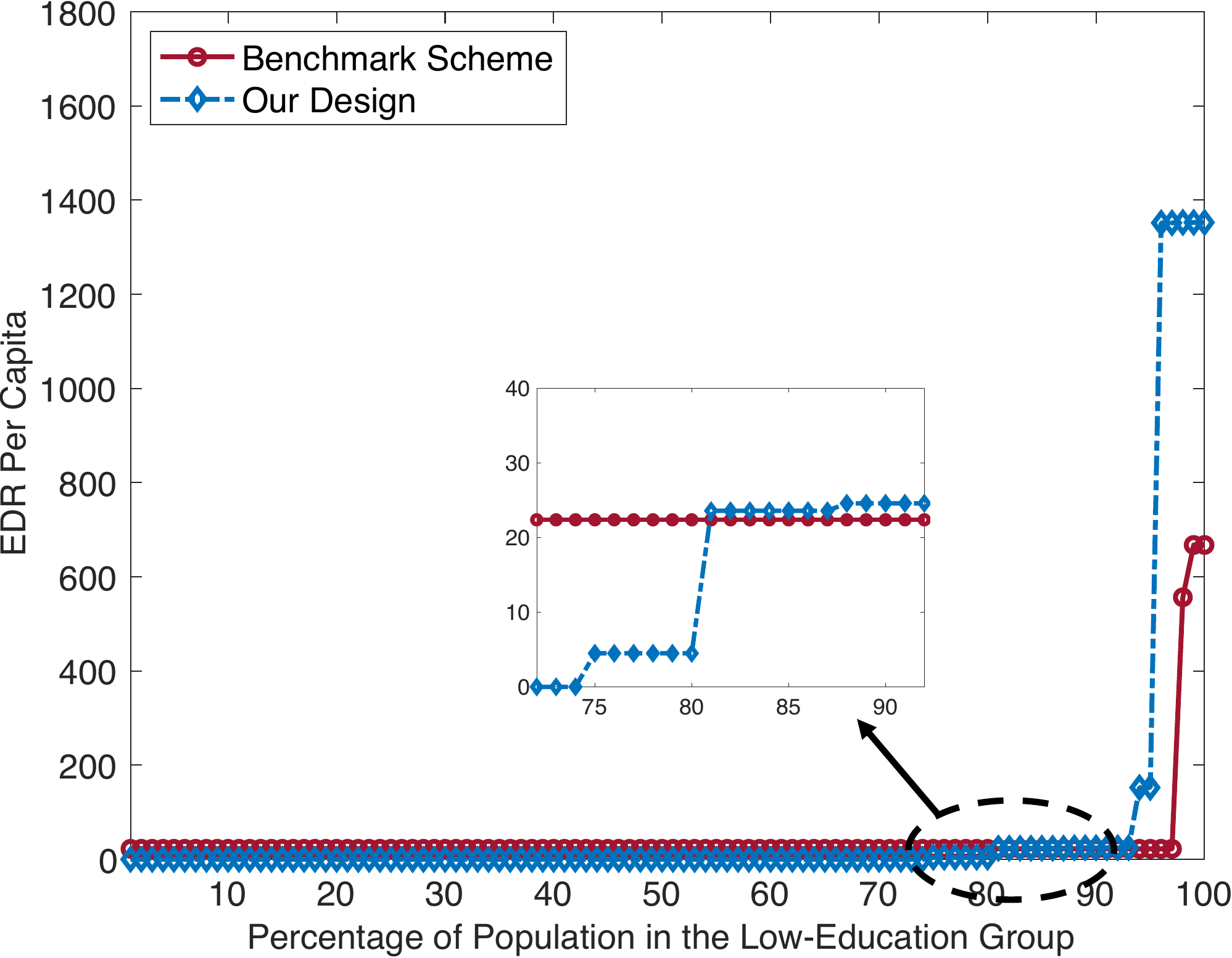}
  \caption{\small EDR in the low-education group.} \label{comp_edu}
\end{subfigure}
   \caption{Performance comparison w.r.t. EDR (Mbps) in various minority groups.} \label{comp_eval}
\end{figure*}

Next, we present the comparison between two designs under the consideration of societal factors. With no bias, we assume that low-income group refers to the one with ``less than \$49,999'' and low-education group refers to the one with ``no higher than high school''. As shown in Fig.\ref{comp_eval}, our design in every scenario outperforms the benchmark scheme in serving under-represented users by 3-10\%, and especially so when it comes to serving the users with high-rate service demands. On the other hand, it is interesting to note that the area under curve (AUC) represents the overall provisioned services (in Mbps). Obviously, our design in all three scenarios are superior than the benchmark scheme w.r.t. the overall service rates.

\section{Discussion}
This work presents a proof-of-concept study on our pioneering attempt on the socio-technological integration. We reveal many insightful findings, yet there are several open problems worthwhile to be explored. This includes, but is not limited to, the following aspects. 

\subsubsection{Other societal factors} In this work, we only present three sociodemographic characteristics, but they cannot fully capture all the contributing factors to a community' communication service needs in the aftermath of disasters. Different subpopulations have their unique needs for communication services. Therefore, future research will include more detailed characterization of the community, such as geographic regions (i.e., rural or urban), health and mobility condition, homeownership (i.e., rent or own), and household with elderly or children, etc. Moreover, the need for communication service is not solely affected by any univariate factor, but instead, it is generally biased by a mix of multiple societal factors. A future work that could possibly unveil such complex correlation will have a profound impact.

\subsubsection{Interdependence of communication services}
In addition to the sociodemographic characteristics that relate to human, post-disaster communication service provisioning should also take human-irrelevant factors into consideration. This could be transportation systems, power-line systems and other communication-interdependent systems. The rationale is that solely restoring the functionality of communication infrastructures does not necessarily mean users' communication services are re-gained, because they may still suffer from power outage that causes the incapability in powering or charging their communication devices at home \cite{esmalian2021susceptibility}. 


\section{Conclusion}
In this work, we studied how sociodemographic factors affect people's perception, susceptibility and demands on communication services in the aftermath of disasters. From there, we optimized post-disaster communication resource allocation with an aim to provide precise service provisioning, thus promoting social equality. On the one hand, we made technical contributions by proposing a novel empirical model and developing solutions to the large-scale optimization problem. On the other hand, this study shows the promise of socio-technological integration in post-disaster communication service restoration for improving the social equality and informing the development of a more equitable and sustainable community. 

\section{Acknowledgement}
The authors would like to thank Dr. Ali Mostafavi from Texas A\&M University for the Hurricane Harvey household survey data support. 

\bibliography{disasterXcom}
\bibliographystyle{IEEEtran}


\begin{IEEEbiographynophoto}{Jianqing Liu}
received the Ph.D. degree from University of Florida in 2018. He is currently an assistant professor in the Department of Electrical and Computer Engineering at University of Alabama in Huntsville. His research interest is to design secure, energy- and spectrum-efficient protocols for various advanced wireless systems. He is the recipient of four best paper awards including the 2018 Best Journal Paper Award from IEEE Technical Committee on Green Communications \& Computing (TCGCC).
\end{IEEEbiographynophoto}
\vspace{-0.1in}

\begin{IEEEbiographynophoto}{Shangjia Dong}
received the Ph.D. degree from Oregon State University in 2018. He is currently an assistant professor in the Department of Civil and Environmental Engineering and a core faculty in Disaster Research Center at University of Delaware. His research interest is to investigate the resilience of coupled human-infrastructure system and derive smart solutions for risk-informed decision-making.
\end{IEEEbiographynophoto}
\vspace{-0.1in}

\begin{IEEEbiographynophoto}{Thomas Morris}
received the Ph.D. and M.S. degree from Southern Methodist University in 2008 and 2001, respectively. He is a full professor and eminent scholar in the Department of Electrical and Computer Engineering at University of Alabama in Huntsville (UAH). Before joining UAH, he worked at Mississippi State University as an associate professor and center director from 2008-2015 and worked as a digital designer at Texas Instruments from 1991-2008. His current research involves security for industrial control systems (SCADA, Smartgrid, Smart meters, process control systems).
\end{IEEEbiographynophoto}

\end{document}